\documentclass[%
 aip,
 amsmath,amssymb,
 reprint,%
]{revtex4-1}

\usepackage{graphicx}
\usepackage{dcolumn}
\usepackage{bm}

\usepackage[utf8]{inputenc}
\usepackage[T1]{fontenc}
\usepackage{mathptmx}
\usepackage{etoolbox}

\makeatletter
\def\@email#1{
 \endgroup
 \patchcmd{\titleblock@produce}
  {\frontmatter@RRAPformat}
  {\frontmatter@RRAPformat{\produce@RRAP{*#1\href{mailto:#1}{lynd@che.utexas.edu}}}\frontmatter@RRAPformat}
  {}{}
}%
\makeatother
\begin{document}


\title[Computational Stochastic Mechanics of a Simple Bound State]{Computational Stochastic Mechanics of a Simple Bound State}

\author{Nathaniel A. Lynd*}
 \affiliation{$^{1}$McKetta Department of Chemical Engineering, The University of Texas at Austin, Austin, Texas, USA}
 
\email{}

\date{\today}

\begin{abstract}
Stochastic mechanics is based on the hypothesis that all matter is subject to universal modified Brownian motion. In this report, we calculated probability density distributions using concepts of stochastic mechanics independent of bootstrapping with a known wave function. We calculate a velocity field directly from the potential and total energy and then use the resultant velocity field to do a stochastic Langevin integration over histories to create the probability density distribution for particle position. Using the harmonic oscillator as a minimally sufficient system for our exposition, we explored the effects of spatial and time discretization on solution noise. We explore the effect of energy defect off of the ground state energy on the velocity field, which dictates how a particle interacts with the background of stochastic fluctuations in position, and describes how repulsive drift (negative defect) and constructive oscillation (positive defect) end a stable state as its basin of stability in the velocity field shrinks with increasing energy defect. The results suggest a pathway for future development of stochastic mechanics as a numerical strategy to describe the structure of physical quantum systems for applications in chemistry, materials and information sciences.
\end{abstract}

\maketitle

\section{\label{sec:intro}Introduction:}

In this report, we establish some basic computational principles for carrying out numerical calculations in stochastic mechanics to produce numerical probability density distributions for indistinguishable particles subject to an arbitrary potential. The numerical aspects are spatial and temporal resolution, collection of statistics, convergence, identification of eigenstate energies, and state lifetime as a function of energy defect, defined as the degree to which a particle’s energy differs from the eigenstate energy. The results of this study set a course for future improvement of numerical stochastic mechanics toward the eventual goal of enabling the emergent prediction of structure and reactivity of quantum systems.

Stochastic mechanics is based on a probabilistic interpretation of quantum mechanics. The earliest developments of a probabilistic interpretation of quantum mechanics were reported by Fürth,\citep{furth,furth_translation} who explored the connections between classical statistics and quantum mechanics. Fényes later proposed that quantum phenomena could be described as a Markov process which underlies the uncertainty principle and that all matter was subject to universal Brownian motion.\citep{fenyes} Nelson developed the concept of stochastic mechanics from the earliest insights by Fürth and Fényes,\citep{nelson1} and later Carlen identified characteristics of stable solutions to Nelson’s stochastic mechanics as conservative diffusions.\citep{carlen} A consequence of stochastic mechanics is that wave-particle duality manifests as a result of universal modified Brownian motion of particles. McClendon and Rabitz demonstrated the stochastic mechanical picture of a two-slit diffraction experiment where averaging over trajectories of particles resulted in an apparent diffraction pattern, but without any explicit wave interference.\citep{mcclendon} Wave-particle duality can be modeled as a consequence of modified Brownian motion rather than constructive or destructive interference between wave functions. Tunneling was framed as being a phenomenon due to the stochasticity of Brownian motion, and the dynamics and correlations within atoms were explored as well, in addition recently by Carosso.\citep{carosso} All of these calculations were based on known wave function solutions to the Schrödinger equation.

Stochastic mechanics  is not in use today as a predictive tool for chemistry or materials science. There are a few possibilities as to why the theory was not developed to numerical implementation. The first is that Kohn and Sham’s landmark paper on density functional theory was published the year prior to Nelson’s foundational article introducing the concept of practical stochastic mechanics.\citep{nelson1,kohn} Another reason is that the approach has been marred by some debate as to its validity. The applicability of stochastic mechanics to quantum mechanics has been proposed previously,\citep{wallstrom,hardy1,garba1} argued against,\citep{gillespie1,gillespie2} and further argued in favor of.\citep{wallstrom,hardy2,garba3} The criticism of stochastic mechanics was generally levied at its seeming incompatibility with a wave function. However, there is no wave function needed in stochastic mechanics. In fact, the lack of a wave function and the centrality of the real-valued probability density distribution is a central postulate of the approach.\citep{furth} In addition, true wave functions are rarely calculated as solutions to the Schrödinger equation for practical prediction of the structure and reactivity of matter by any modern method.\citep{simons} Therefore, I believe the criticism of stochastic mechanics on the grounds offered up by others and even Nelson himself to be premature.\citep{gillespie1,gillespie2,nelson2} In the opinion of the author, the right question to ask is if stochastic mechanics can be developed to model real experimental systems and make meaningful predictions and not if it is formally equivalent to the theory of quantum mechanics (or any theory) in every aspect. But before the experimental relevance of stochastic mechanics can even be investigated, the numerical characteristics of the theory must mature. A further headwind for adoption of numerical stochastic mechanics is the heavily computational nature of the likely implementation. The potential algorithm would have been too computationally demanding for meaningful collection of statistics on serial processors available during the time that stochastic mechanics was first conceptualized by Fürth and eventually Nelson.\citep{furth,nelson1} However, given the high parallelism available today on modern central processing units, graphical processing units, and accompanying numerical libraries, the opportunity is now imminent to develop stochastic mechanics as a general and predictive method of calculating the structure and reactivity of matter. In should be stated that we are not proposing to develop stochastic mechanics as a replacement to the formalism enshrined in quantum mechanics, rather to advance the predictive utility of stochastic mechanics for chemistry, materials, and information science applications, and to explore the consequences, if any, of the perspective advanced by stochastic mechanics.

In stochastic mechanics, the fundamental motion is a Markovian diffusion process described by the Langevin stochastic differential equation for the location coordinate $q(t)$ of a particle with mass $m$. 
\begin{equation}
q(t+\delta t) = q(t) + v(q) \delta t + \xi \sqrt{2 D \delta t}
\label{eq:langevin}
\end{equation}
with diffusion coefficient
\begin{equation}
D = \frac{\hbar}{2 m}
\label{eq:diffusioncoefficient}
\end{equation}
and time increment $\delta t$ and a normally distributed random real number with unit variance $\xi$ and average of zero. Lighter particles are more affected by the random fluctuation term. In an analogy with Brownian diffusion, $v(q)$ is referred to as the forward drift or the velocity field with units of velocity.\citep{garba2} The $v(q)$ can be calculated from a known solution to the Schrödinger equation given by the wave function $\psi(q)$ for the system
\begin{equation}
v(q,t) = \frac{\hbar}{m}\left\{ \Re{\left[\frac{\nabla\psi}{\psi}\right]} + \Im{\left[\frac{\nabla\psi}{\psi}\right]} \right\}
\label{eq:velocityfield}
\end{equation}
Where $\Re{}$ refers to the real and $\Im{}$ the imaginary components of $\psi$. $v(q,t)$ contains the structure of the system and describes how the particle location combines with random fluctuations with magnitude given by $\xi (2 D \delta t)^{1/2}$. In this article, we refer to $v(q,t)$ as the velocity field, and we first must decouple the description of $v(q,t)$ from the wave function in \ref{eq:velocityfield}, and explore methods of calculating $v(q,t)$ directly from the interaction potential, $V$, and energy of a single particle, $E$. This can be accomplished: By taking the gradient of equation \ref{eq:velocityfield}, using the definition of $v(q,t)$ \ref{eq:velocityfield}, and together with the time-independent Schrödinger equation, generalizing to three-dimensions, then a new equation relating the interaction potential $V$ and the energy $E$ to the time-independent velocity field $v(r)$ for single particle system can be written
\begin{equation}
-\frac{\hbar}{2}\left(\nabla \cdot v \right)-\frac{1}{2}m \left(v\cdot v \right) + V = E
\label{eq:bridge}
\end{equation}
Equation \ref{eq:bridge} has been derived \textit{via} a different approach by Nelson.\citep{nelson1} Equation \ref{eq:bridge} is the conceptual backbone of time-independent computational stochastic mechanics. For a given energy $E$ and potential $V$, a velocity field $v$ is defined by equation \ref{eq:bridge} that describes how the stochastic fluctuations and interactions combine to create statistics of particle position using the Langevin integration of equation \ref{eq:langevin}. Without the fluctuations, the particle would drift to the minimum of $V(x)$ (defined here as $x = 0$) with its motion described by the velocity field $v(x)$ where $v(0) = 0$. With the velocity field for a given $V$ and $E$, $v(r)$, an ensemble of particle locations are generated stochastically according to equation \ref{eq:langevin} and the particle locations can be integrated over time until a probability density at a desired noise is resolved. This prescription presents many numerical questions: The error due to discretization of time, $\delta t$, the reduction of solution noise with iteration at different time discretization, the effect of spatial resolution $\delta q$ and $n$, and the effect of energy defect, $\delta E$, if one were searching for an unknown stable solution. We will also explore the changes in the velocity field with $\delta E$ and the consequences for stability. There are very few exactly solvable quantum mechanical systems, and so in this first work, we use the harmonic oscillator as a minimally sufficient basis for our numerical and analytical exposition of basic concepts of computational stochastic mechanics.

\section{\label{sec:numerics}Analytical and Numerical Methods:}
The quantum harmonic oscillator can be modeled as a particle subject to random fluctuations in position that are countered by the influence of the velocity field unique to the ground state given by equation \ref{eq:HOfield}
\begin{equation}
v(x) = -\left(\frac{k}{m}\right)^{1/2} x
\label{eq:HOfield}
\end{equation}
Inserting the harmonic oscillator velocity field \ref{eq:HOfield} and the classical harmonic potential $(1/2) k x^2$ ($k$ force constant, $x$ displacement) into \ref{eq:bridge}, and solving for energy recovers the exact ground state energy for the quantum harmonic oscillator.
\begin{equation}
E_0 = \frac{\hbar}{2} \left(\frac{k}{m}\right)^{1/2}
\label{eq:HOenergy}
\end{equation}
The stochastic mechanical picture of the harmonic oscillator is numerically consistent with the analytical quantum mechanical picture. With a given potential, a particle’s energy determines its velocity field, which for certain values of the energy results in a diffusion that is stable for the duration of a simulation ($\tau_{max}$). For a stochastic simulation, the starting location of a particle can be initiated at any location within the computational domain $[-L,+L]$, and after sufficient Langevin integration using equation \ref{eq:langevin}, the distribution can be normalized and should be identical to the equivalent squared ground state wave function for the quantum harmonic oscillator representing the probability density distribution of the particle $\left| \psi \right|^2$. For other values of the energy, which we describe as an energy defect off of the ground state energy, $\delta E$, the velocity field can be solved as a numerical initial value problem starting from the minimum of the classical potential at $x = 0$. We define collocation arrays for space, $x_j$, and velocity field $v_j$. The collocation array for $x$ spans $x_0 = -L$ and $x_{n-1} = +L$ with $n$ collocation points, and the collocation for $v_j$ represents the value of the velocity field at each corresponding point $x_j$. The velocity field can be solved using a finite difference approximation of equation \ref{eq:bridge}. The following finite difference expressions for incrementing $j$ from the initial condition $v(x_{n/2} = 0) = v_{n/2} = 0$ to yield the solution on $[0,+L]$
\begin{equation}
v_{j+1} = v_j - \frac{\delta x}{\hbar}\left[m v_j^2 - k x_j^2 + 2 \left(E_0 + \delta E \right) \right]
\label{eq:inc}
\end{equation}
And decrementing $j$ to yield the solution on $[0, -L]$
\begin{equation}
v_{j-1} = v_j + \frac{\delta x}{\hbar}\left[m v_j^2 - k x_j^2 + 2 \left(E_0 + \delta E \right) \right]
\label{eq:dec}
\end{equation}
With the velocity field defined at discrete collocation points, a polynomial interpolant was created from the collocation array to create a smooth and continuous function that can be evaluated at arbitrary $x$ during Langevin integration of observations (\textit{i.e.}, sum-over-histories) using equation \ref{eq:langevin}, which is used to update particle position at every iteration incrementing the simulation time by $\delta t$. For convergence analysis, the entire collocation array $v_j$ was used to form the interpolant. For analysis of lifetime with non-zero energy defect, only the three bounding collocation points of the current particle position were used.

An array data structure was used to keep track of the sum-over-histories of particle location. The array represents discretized space into bins where $\rho_j$ represents the fraction of the number of time steps where the particle is present in the interval between $x_j$ and $x_j + \delta x = x_{j+1}$. At each iteration, $\rho_j$ is incremented by one at $j$ for the particle located within $[x_j, x_{j+1}]$. At the end of the calculation, the $\rho_j$ is normalized to unity on $[-L,+L]$ using either Simpson’s method or the trapezoidal method if the number of collocation points are respectively odd or even. For presentation of results and calculation of solution noise relative to the known probability density distribution $\left|\psi \right|^2$, the distribution $\rho_j$ is shifted by $\delta x/2$ in $x$ so that $x_j$ describes the center of each array element. Solution noise $\sigma_n$ was estimated on a per-collocation-point basis by the squared difference between the known probability density distribution (shifted by $\delta x/2$) and $\rho_j$ normalized by the number of collocation points $n^{-1}$ and collected at logarithmically spaced iterations:
\begin{equation}
\sigma_n = \frac{1}{n} \sum_{j=0}^{n-1} \left\{\left[\psi(x_j + \delta x/2) \right]^2-\rho_j \right\}^2
\label{eq:noise}
\end{equation}
For simulations with $\delta E \neq 0$, \textit{i.e.}, an energy defect off of the ground state, the lifetime of a simulation was estimated as the time at which the particle coordinate drifted a distance of $5L$, $\pm\infty$, or a non-numerical value. The lifetime of a simulation with non-zero energy defect was typically very short, with immediate ejection of the particle far from the equilibrium position near $x = 0$ to outside the simulation box at $\pm L$.

\section{\label{sec:results}Results and Discussion:}
\begin{figure}
\includegraphics[scale=0.80]{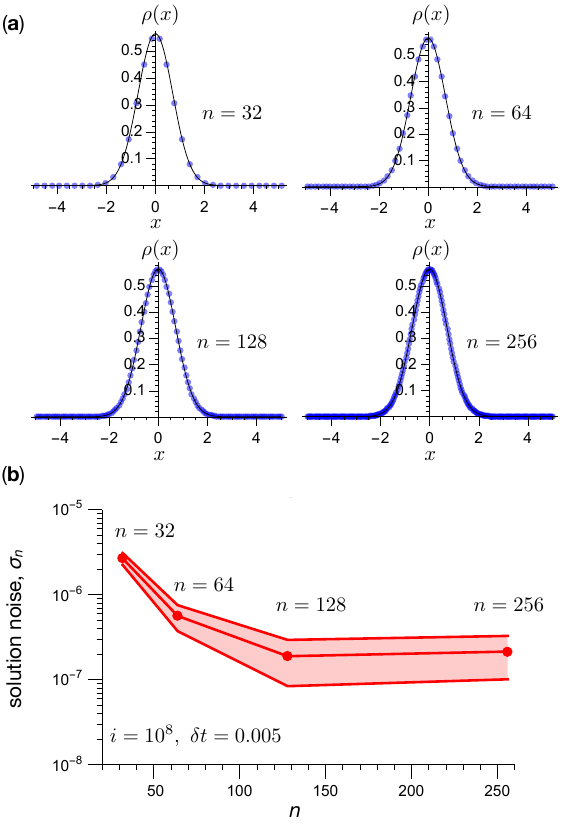}
\caption{\label{fig:1} (a) Normalized residency histograms (probability densities) after $10^8$ iterations of $\delta t = 0.005$ with $n = 32$, $64$, $128$, and $256$ collocation points. (b) Solution noise measured using equation \ref{eq:noise} of the states shown in (a). The points are the mean and shaded area represents the standard deviation of 12 simulations.}
\end{figure}

The numerical methods described above were implemented in C++ using the Eigen library for vector and matrix data structures and operations. The Mersenne Twister was used for generation of pseudo-random numbers and the mt19937 object was seeded with the system time. As such, computations were run such that shorter duration calculations were interspersed with longer duration calculations to avoid artifacts due to identical pseudo-random number generator seeding. A preliminary investigation of convergence to a low-noise solution of the harmonic oscillator for various spatial and time discretization was explored. Results of this preliminary investigation are shown in Figure \ref{fig:1}. As seen in Figure \ref{fig:1}a, the residency histogram data points at $n = 32$, $64$, $128$, and $256$ are plotted atop the probability density of the ground state quantum harmonic oscillator. At all $n$, the residency histogram agreed to high accuracy with the ground state probability density with a weak relationship between solution noise calculated using equation \r ef{eq:noise} and n shown in \ref{fig:1}b. Significantly, it was found that for the one-dimensional harmonic oscillator that $n = 128$ represented an apparent minimum in solution noise after $10^8$ iterations with $\delta t = 0.005$, which was generally sufficient for convergence to low-noise solutions. There was a weak dependence of solution noise on the spatial resolution of the residency histogram. We selected $n = 128$ for all subsequent calculations. 

\begin{figure}
\includegraphics[scale=0.75]{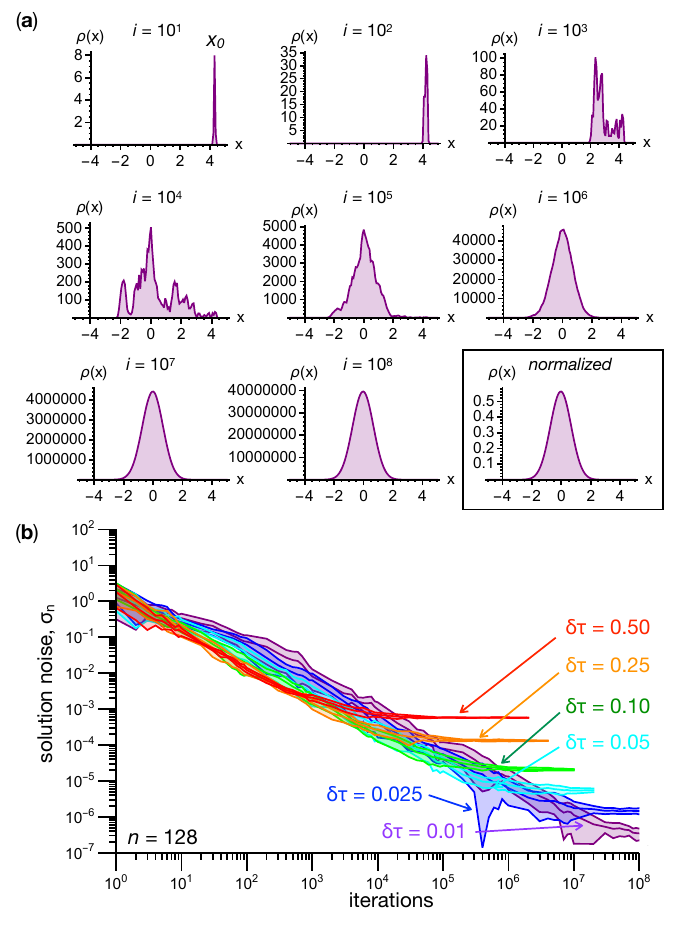}
\caption{\label{fig:2} (a) Cumulative solution histograms for $n = 128$ and $\delta t = 0.001$ at iteration $i$ and normalized after $10^8$ iterations. (b) Solution noise was calculated as a function of iteration count for different time step sizes demonstrating that convergence to low-noise solutions may be accelerated using dynamically-sized $\delta t$ which starts large at high noise and decreases with iteration count. The solid line is the mean and shaded region represents the standard deviation across 12 repeated simulations.}
\end{figure}

Numerical calculations using stochastic mechanics as we are doing here have not been previously done to the best of our knowledge, and so an evolution of the residency histogram of a stochastic mechanic simulation is shown in Figure \ref{fig:2}a. The initial particle position was selected randomly as $x_0$ within $[-L,+L]$, and then updated iteratively according to equation \ref{eq:langevin} with velocity field described as the solution to equation \ref{eq:bridge}. The evolution of the residency histogram is shown at decades of iterations up to $10^8$ iterations after which the residency histogram was normalized to the probability density distribution of the sum-over-history of the simulation. Probability density distributions calculated using both stochastic mechanical and quantum mechanical approaches were numerically indistinguishable.

\begin{figure}
\includegraphics[scale=0.75]{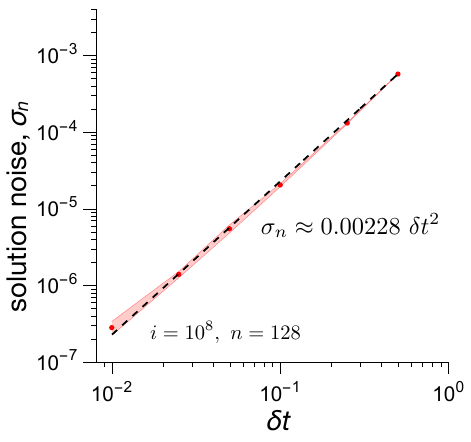}
\caption{\label{fig:3} The relationship between converged solution noise at 108 iterations at $n = 128$ as a function of time increment $\delta t$. Points are the averages across 12 simulations whereas the shaded area represents the standard deviation. The dashed line represents a power-law fit to the averaged data of $0.00228 \delta t^2$.}
\end{figure}

The discretization of time introduces numerical error. Convergence to a low-noise solution was investigated as a function of time-increment across the domain $\delta t = 0.001-0.500$ in Figure \ref{fig:2}b. The mean solution noise is indicated by the curve bounded by the shaded region representing the standard deviation across 12 differently seeded simulations. Both the rate of decrease of solution noise with iteration and the ultimate converged solution noise depended on $\delta t$. Solution noise increased at early iterations most rapidly with a large $\delta t$ of 0.500. However, the solution noise converged to \textit{ca.} $10^{-3}$ as measured using equation \ref{eq:noise}. The larger $\delta t$ leads to larger updates in position that create a more fluctuating probability density distribution due to overly large particle displacements that prevent further decreases in solution noise. Simulations with larger $\delta t$ converge more quickly but to higher noise solutions. Significantly, the smallest $\delta t$ produced the slowest initial decrease in noise but ultimately the lowest solution noise after the full $10^8$ iterations. These results suggest that calculations in stochastic mechanics may benefit from dynamically-sized $\delta t$ with larger $\delta t$ taken early on for rapid reduction of error followed by smaller $\delta t$ for solution refinement. Such a method will be explored in future work. The relationship between ultimate solution noise and $\delta t$ is shown in Figure \ref{fig:3}. The solution noise scaled approximately with the square of $\delta t$ when fit to a power law. The above calculations established some preliminary characteristics of stochastic simulations at exact ground state energy. We next explore the effect of energy defect on a stochastic calculation of state, which is a situation that may arise in the investigation of a new system with unknown ground state energy.

\begin{figure}
\includegraphics[scale=0.8]{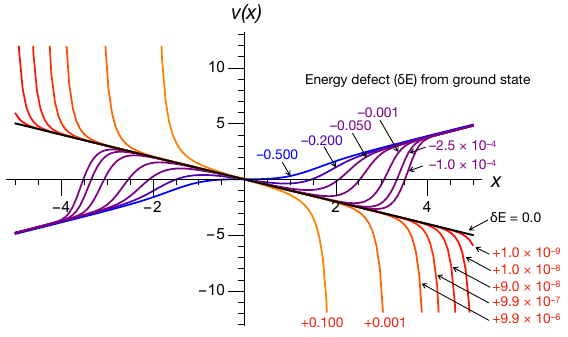}
\caption{\label{fig:4} The effect of energy defect ($\delta E$) on the velocity field changes the characteristic of the particle motion.}
\end{figure}

The effect of energy defect can be incorporated into a stochastic simulation by solving equation \ref{eq:bridge} numerically for the velocity field at a defective energy state, \textit{i.e.}, $E + \delta E$. This type of operation would be done in order to search for stable states of uncharacterized potentials. The explicit integrators equations \ref{eq:inc} and \ref{eq:dec} were used to solve for discrete collocation arrays in $x_j$ and $v_j$ and only piecewise interpolation using a quadratic function at arbitrary $x$ was used from the resultant discrete collocation arrays in $x_j$ and $v_j$ to evaluate arbitrary $v(x)$ where $x_{j-1} < x < x_{j+1}$. A series of velocity fields at negative, positive, and zero $\delta E$ are shown in Figure \ref{fig:4}. The $\delta E = 0$ state is represented by the solid black line with negative slope, described analytically by equation \ref{eq:HOfield} but the data in Figure \ref{fig:4} were calculated numerically using equations \ref{eq:inc} and \ref{eq:dec}. Negative and positive $\delta E$ affect $v(x)$ differently, which in turn affects the nature of the particle’s motion within the potential when combined with the stochastic fluctuations. The smaller the magnitude of $\delta E$, the defective velocity function follows the ground state velocity function over a larger domain in $x$. For negative $\delta E$, first at large $\pm L$ the slope of the velocity function changes sign. As $\delta E$ becomes increasingly negative, more of $v(x)$ takes on a positive slope which would increase the likelihood of a particle exhibiting repulsive drift away from the minimum of the potential located at $x = 0$. In that instance, it would be very unlikely for a stochastic fluctuation to return the particle to a location within the basin of stability of the velocity field, \textit{i.e.}, the portion of $v(x)$ that is coincident with the ground state velocity field. For positive $\delta E$, divergences appear at the fringes of the velocity field and would manifest different behavior. With increasing $\delta E$, the positive and negative divergences encroach toward the center and reduce the size of the basin of stability. Once the particle fluctuates near a divergence, it would undergo constructive oscillation with increasing amplitude until it reached $\pm\infty$. Small $\delta E$ could result in a stable, long-lived state. Next, we investigated how $\delta E$ affects the lifetime of a state ($\tau$).

\begin{figure}
\includegraphics[scale=0.8]{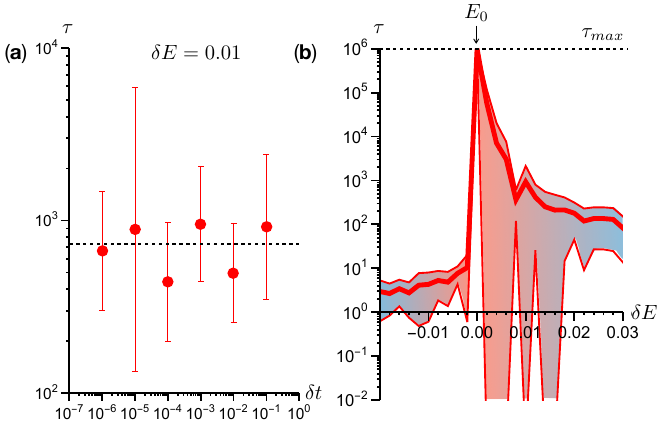}
\caption{\label{fig:5} (a) Stability lifetime ($\tau$) as a function of simulation time increment ($\delta t$). (b) $\tau$ as a function of $\delta E$ a precipitous spike in stability at $\delta E = 0$.}
\end{figure}

The state lifetime $\tau$ was investigated with a series of calculations with a set maximum lifetime of $\tau_{max} = 10^6$ atomic time units. To estimate the state lifetime $\tau$, rather than placing the initial particle position anywhere within the simulation domain, we placed it pseudo randomly within 10\% of the minimum of the potential to maximize the likelihood of the particle first being placed in the basin of stability of the velocity field. First, we investigated if variation in time increment $\delta t$ could produce an artificially long or short stability lifetime $\tau$ at a fixed energy defect of $\delta E = 0.01$ and $\delta t$ was varied from $10^{-6}$ to $10^{-1}$ and shown in Figure \ref{fig:5}a. The calculations were averaged over ten simulations with the error bars representing the standard deviation. The dashed line indicates the average across the domain of $\delta t$. The average $\tau$ at each $\delta t$ were all within the standard deviation of their combined average. Thus, variation in $\delta t$ from $10^{-6}$ to $10^{-1}$ did not appear to produce an artifact of stability, and $\delta t = 10^{-3}$ was selected to measure the effect of $\delta E$ on $\tau$, which is shown in Figure \ref{fig:5}b. The data in Figure \ref{fig:5}b cover the domain from $\delta E = -0.02$ to $+0.03$ and are averaged across ten simulations with the bold curve representing the average and the shaded region the standard deviation of the averaged lifetimes at each $\delta E$. Negative $\delta E$ was characterized by the minimum $\tau$, which drifted slightly upward as $\delta E = 0$ was approached. At $\delta E = 0$ an abrupt spike to $\tau = \tau_{max}$ occurred, with no deviation across the ten simulations. Further increases in $\delta E$ to positive values caused a precipitous decrease in $\tau$ with a large standard deviation. While the divergences in the velocity field at positive $\delta E$ were abrupt, the relatively wider basin of stability allows for longer $\tau$, which decreased as $\delta E$ was further increased. Presumably, as $\delta E$ were to be increased further, eventually a longer-lived excited state would be found, with an additional spike in stability. For an arbitrary potential, the calculation of the velocity field of a longer-lived excited state $v_{ex}(x)$ would need to be built upon a prior calculation of the ground state velocity field $v_0(x)$ so that an appropriate initial or boundary value can be identified for the numerical solution of $v_{ex}(x)$ using equation (4) with $v_{ex}(L \gg 0) = v_0(L \gg 0)$ and $v_{ex}(-L \ll 0) = v_0(-L \ll 0)$ as initial values. Excited states may contain divergences in the velocity fields that will presumably produce minima in the probability density. A detailed exploration of properties of excited states is beyond the scope of this current work.

\section{\label{sec:conclusion}Conclusions and Outlook:}
Stochastic mechanics is based on the hypothesis that all matter is subject to universal Brownian motion and that probability density distributions for particle location can be calculated either by solution of the Fokker-Planck equation or as we have done here by Langevin integration to a low-noise probability density. For the first time, we have calculated probability density distributions using concepts of stochastic mechanics. An advantage of the stochastic mechanical simulation approach is that it allows the calculation of a three-dimensional probability density from a classical potential without bootstrapping by a previously known wave function. The numerical exposition here presented a number of new concepts, and highlighted the need for future development. The numerical solution of the bridge equation at a given energy provides access to a velocity field that describes how a particle interacts with a background of stochastic fluctuations in its position. Intermediate and final solution noise was found to depend on the time increment used for Langevin integration, which suggests that numerical strategies for dynamically sizing the time increment may accelerate convergence. The energy determines the size of the basin of stability for the diffusion, with the basin of stability extending over all space at the ground state energy. With energy defect, the stable state can decay either by repulsive drift for negative defect or constructive oscillation for positive defect. Unstable state lifetime depends on its energy relative to the ground state energy. Methodology will need to be developed to search for stable states for arbitrary potentials and energies, as well as excited states. Finally, the numerical strategy is highly parallelizable.

\begin{acknowledgments}
NAL thanks the Cockrell School of Engineering at the University of Texas at Austin for sabbatical leave to focus on developing the concepts in this article. NAL also acknowledges the Welch Foundation (grant no. F-2201-20240404) for its generous support of this research.
\end{acknowledgments}

%

\section*{References}

\nocite{*}
\bibliography{aipsamp}

\end{document}